\documentclass[aps,twocolumn,nofootinbib,showpacs,preprintnumbers,amsmath,amssymb]{revtex4}%    % Enable this line and disable the
\usepackage{graphics}
\usepackage{graphicx}
\usepackage{dcolumn}% Align table columns on decimal point
\usepackage{bm}% bold math
\newcommand{\be}{\begin{eqnarray}}
\newcommand{\ee}{\end{eqnarray}}

 \newcommand{\gsim}{\mathrel{\hbox{\rlap{\lower.55ex \hbox {$\sim$}}
                   \kern-.3em \raise.4ex \hbox{$>$}}}}
\newcommand{\lsim}{\mathrel{\hbox{\rlap{\lower.55ex \hbox {$\sim$}}
                   \kern-.3em \raise.4ex \hbox{$<$}}}}

\newcommand \beq{\begin{equation}}
\newcommand \eeq{\end{equation}}
\newcommand \bea{\begin{eqnarray}}
\newcommand \eea{\end{eqnarray}}

\newcommand \br{{\mathbf r}}

\begin{document}

\title{Static $\bar Q Q$ Potentials
and the Magnetic Component of QCD Plasma near $T_c$}

\author{Jinfeng Liao$^{1}$}\email{jliao@lbl.gov}
\author{Edward Shuryak$^2$}\email{shuryak@tonic.physics.sunysb.edu}
\affiliation{$^1$Nuclear Science Division, Lawrence Berkeley National
Laboratory, Berkeley, CA 94720, USA.\\
$^2$Department of Physics and Astronomy, State
University of New York, Stony Brook, NY 11794, USA.}


\begin{abstract}
Static quark-anti-quark potential encodes important information on the
chromodynamical interaction between color charges, and recent lattice
results show its very nontrivial behavior near the deconfinement temperature $T_c$.
In this paper we study such potential in the framework of the
``magnetic scenario'' for the near Tc QCD plasma, and particularly focus on the linear part
(as quantified by its slope, the tension) in the potential as well as the strong splitting between the free energy and internal energy.
By using an analytic ``ellipsoidal bag'' model, we will quantitatively relate the free energy tension to the magnetic condensate density and relate the internal energy tension to the thermal monopole density. By converting the lattice results for static potential into density for thermal monopoles we find the density to be very large around Tc and indicate at quantum coherence, in good agreement with direct lattice calculation of such density. A few important consequences for heavy ion collisions phenomenology will also be discussed.

\end{abstract}

\maketitle

\section{Introduction}

\begin{figure}
\begin{center}
\includegraphics[width=7.5cm]{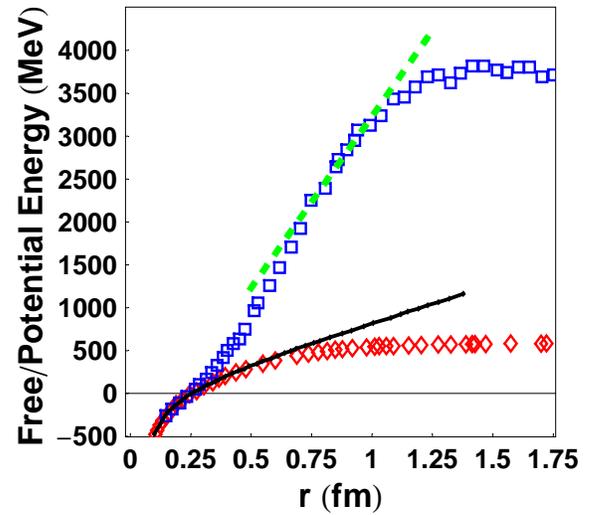}
\end{center}
\caption{\label{fig_splitting} The static $\bar{Q}Q$ potential at $T\approx T_c$ (adapted from \cite{Kaczmarek_Zantow}). The blue boxes are for the internal energy $V(r)$ while the red diamonds are for free energy $F(r)$,  with the green dashed line indicating the strong linear rise in $V(r)$ for $r\in (0.5,1)\rm fm$ and the black solid line showing the vacuum $\bar{Q}Q$ potential. }
\end{figure}

The interaction potential between static quark and anti-quark pair
is a traditional observable to study the quark confinement mechanism
in QCD. It was originally
inferred from heavy meson spectrum and Regge trajectories, and
has then been extensively studied in lattice gauge theories, for reviews see e.g.
\cite{Bali_review,Greensite}. Its vacuum ($T=0$) form is well known,
usually represented as a sum of a Coulomb part $V\sim 1/r$,
dominant at small separation between $\bar{Q} Q$, and a linear part $V=\sigma r$
dominant at large separation (see the black solid curve in Fig.\ref{fig_splitting}). The latter implies the confinement of quarks and has been interpreted in terms of chromo-electric flux tube (or ``string'') formation
between well-separated $\bar{Q} Q$ pair. The so-called string tension $\sigma$ in the vacuum ($T=0$) has been consistently
determined by different methods to be
\begin{equation} \sigma_{vac} \approx (426MeV)^2 \approx 0.92\,GeV/fm
\end{equation}

With current RHIC and future LHC experimental programs exploring
excited hadronic matter and quark-gluon plasma (QGP) at increasing
temperature $T$, it is very important to know the finite $T$ form of the static $\bar Q Q$ potential, which has recently been calculated by means of the lattice QCD, see e.g.  \cite{Kaczmarek_Zantow,Kaczmarek_pure_gauge,Petreczky_3f}. At finite temperature, there are actually two potentials associated with a $\bar Q Q$ pair separated by distance $r$: one is the free energy $F(T,r)$ and the other is the internal energy $V(T,r)$, with the difference related to the entropy generated in the medium by the $\bar Q Q$ pair, i.e.
\begin{equation}
V(T,r)=F(T,r)-T\left(\partial F / \partial T\right)=F(T,r)+TS(T,r)
\end{equation}
What is directly evaluated on lattice is the free energy $F(T,r)$ from which the
corresponding $V(T,r)$ and $S(T,r)$ can be inferred \cite{Kaczmarek_Zantow}. While at $T=0$ there is no entropy and the free and internal energies are identical, splitting between the two shall be expected at $T>0$ and may carry key information about the medium and deconfinement transition near $T_c$.

\begin{figure}
\begin{center}
\includegraphics[width=7.5cm]{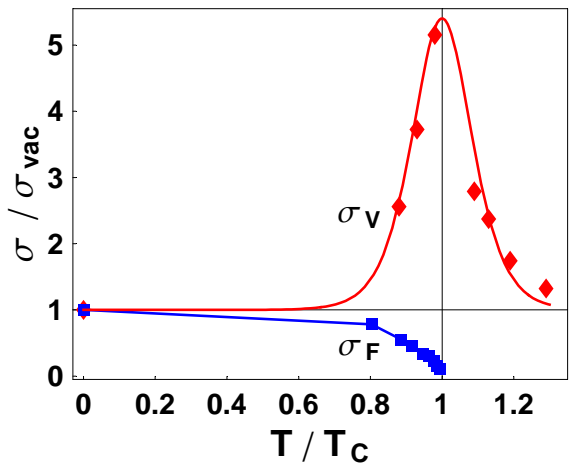}
\end{center}
\caption{\label{fig_tension} Effective string tensions in the free
energy $\sigma_F(T)$ (from \cite{Kaczmarek_pure_gauge}) and the
internal energy $\sigma_V(T)$ (extracted from
\cite{Kaczmarek_Zantow}). }
\end{figure}

The lattice results indeed show remarkably different  potentials $F(T,r)$  and  $V(T,r)$  near $T_c$ (see e.g. Fig.1-4 in
\cite{Kaczmarek_proceedings} and also here Fig.\ref{fig_splitting} adapted from \cite{Kaczmarek_Zantow}). In particular let us emphasize two important points.\\
(i) The tensions (slopes of the potentials at $r$ about $0.3-1\rm fm$) have very different temperature dependeces:  while the tension of the free energy $\sigma_F$ decreases with $T$,  to near zero at $T_c$ (an expected signal of deconfinement), the tension of the internal energy $\sigma_V$ remains nonzero till about $T=1.3T_c$, with a peak value at $T_c$ about 5 times (!) the
vacuum tension $\sigma_{vac}$ (see Fig.\ref{fig_tension}). \\
 (ii) This  drastically different behavior persists to very large distances, where  linear behavior changes to saturated values.
Near $T_c$ the internal energy flattens to huge asymptotic value at large $r\rightarrow\infty$, e.g.
$V(T,\infty)\sim 4\, GeV$  at $T_c$ with the corresponding
 entropy $S(T_c,\infty)\approx 20$
implying huge number of states involved, $\sim exp(20)$.\\
These  features indicate strikingly strong interaction between the static color charges and the medium near $T_c$, which persists into the deconfined phase.

Such static $\bar Q Q$ potentials at finite $T$ are closely connected with a number of phenomenological issues. For example, the consequence of these features for the survival of quarkonium in deconfined plasma is much debated, e.g. on what/which potential should be used \cite{Young:2008he,Zhao:2007hh,Mocsy:2007yj}.
If, as suggested in \cite{bound_states,LS_bound}, the  internal energy is used, $J/\psi$ state would exist even in the deconfined plasma in $1-2T_c$.  Persistence of some baryonic states above $T_c$ is also indicated by other observable like the baryonic susceptibilities\cite{susceptibilities,LS_susceptibilities}.
These potentials also imply significant interaction energy in the quark-gluon plasma and in the many body context this may lead to a large classical plasma parameter $\Gamma$ (defined as the ratio of average interaction energy to average kinetic energy): indeed the $\Gamma$ value in sQGP has been estimated to be above one (about 3) and thus in a typical liquid regime (see for example \cite{Boris_Edward_Ismail,LS_monopole,Ratti:2008jz}). If so, QGP would be a strongly coupled Coulombic liquid, in agreement
with the strong collective flow observed at RHIC, see more in reviews
\cite{sQGP_review,sQGP_ES}.
Apart from QGP phenomenology,  it is  important  to understand the microscopic origin of the potentials, especially the strong splitting between two potentials and the large energy/entropy associated with the static
$\bar Q Q$ pair near $T_c$. Earlier attempts can be found in e.g. \cite{Antonov:2006wz,Megias,Beraudo:2007ky}.

In this paper we will specifically focus on the ``tensions'' $\sigma_F$
and $\sigma_V$ (as shown in Fig.\ref{fig_tension}) related to the linear part of the potentials (while leaving the discussion of
``screening'' behavior at very large distances to further studies). We will provide an explanation in the framework of
the {\it magnetic scenario} of QCD plasma near $T_c$ \cite{LS_monopole,Liao:2008jg,lattice_mag,lattice_mag_density,lattice_mag_eos}. In such a scenario, the near $T_c$ QCD plasma is strongly influenced by the magnetic component, made of relatively light and abundant {\em chromo-magnetic monopoles}.
 Those are quasiparticles above $T_c$ which undergo the Bose-Einstein condensation (BEC) below $T_c$,  enforcing  color confinement  (for reviews see e.g. \cite{Shuryak:2008pz,Liao:2008wb}). 
  Two key points of the present model for the potentials are: First, we identify the $\bar Q Q$ free energy as been probed by an adiabatically ``slow separation'' process while the internal energy by a ``fast separation'' process. Second, we further relate the linear part of potentials with the flux tube formation, enabled by condensed monopoles below $T_c$ while thermal monopoles above $T_c$, between the $\bar Q Q$ pair during the separation process \cite{Liao:2007mj}, and relate the free/internal energy tensions with the condensed/thermal monopoles respectively. These ideas will be elaborated more in Section II and III.

The rest of the paper is structured as follows. In Section-IV we will develop
an analytic ``elliptic flux bag'' model for a static charge-anti-charge pair by solving
the Laplace equation for electric field inside it. This
allows to get the potentials correctly interpolating between
Coulomb at short distance and linear behavior at larger distance. The model will then be
used in section-V to determine the free and potential energies
and relate the extracted $\sigma_F(T)$ and $\sigma_V(T)$ with the
monopole condensate  and the thermal monopole density,
respectively.  
Finally we summarize the results in Section-VI.

\section{ Free v.s. Internal Energy and Slow v.s. Fast Separation}

Let's start by examining the difference between the free energy
and the internal energy. We already introduced the effective string
tensions $\sigma_F(T)$ and $\sigma_V(T)$ as the slopes of linear
parts in $F(T,r)$ and $V(T,r)$ respectively, and emphasized their
quite different $T$-dependencies shown in
Fig.\ref{fig_tension}. While $\sigma_F$ vanishes at $T>T_c$,
$\sigma_V$ survives to at least $1.3T_c$. While $\sigma_F$
monotonously decreases with $T$, $\sigma_V$ peaks at $T_c$ to a
maximal value of 5 times the vacuum string tension $\sigma_{vac}$.
What is the difference in the meaning of $F$ and $V$, and why do
they have such different $T$-dependence? As has been emphasized in \cite{bound_states},
the free and internal energies actually correspond to slow and fast
(relative) motion of the charges, respectively. Let us explain this idea in
more details.

 Consider the ``level crossing''
phenomena, occurring while the separation between charges is changed.
Suppose a pair of static charges (held by external ``hands'') are
moved apart in thermal medium at certain speed $v=\dot L$. 
 For each fixed $L$, there are multiple
configurations of the medium  
populated thermally. When  $L$ is changed, the energies of these configurations are crossing each
other, and at each level crossing there is certain probability
to change between the levels, %  to another which on the long time scale would reach
%the new thermal population. It is precisely such probability for level
%crossing that
 depending on the speed of separation $v=\dot L$. If the motion
is  adiabatically slow, then all the level crossing processes happen with probability $1$: in thermodynamical context this leads to maintained equilibrium
 and maximal entropy/heat generation. If however the pair is separated very fast, then  the level crossing is suppressed  and the medium is no longer in equilibrium
 with the pair. The amount of
  entropy generated is less than in the adiabatic case. In the extreme case one may expect that the pair, if  moving on a time scale much much shorter than the medium relaxation time scale, decouples from the media and produce negligible entropy. It is plausible, therefore, to identify the adiabatic limit as probing the free energy $F(T,L)$ measured on the lattice with the presence of static $\bar Q Q$ pair. The ``internal energy'' $V(T,L)$, on the other hand, is different from $F(T,L)$ by subtracting the entropy term and thus can be probed in the extremely fast limit in which possible transitions among multiple states via
 level crossing do $not$ occur and no entropy is generated.

We emphasize that such phenomenon in thermal medium is a direct analogue of what exists in
pure quantum mechanical context. Perhaps the oldest example is the so called Landau-Zener phenomenon \cite{Landau_QM,Landau_Zener}
of electron dynamics during the vibrational motion of two nuclei
in a diatomic molecule. Specific electron quantum states
$\psi_n(L)$ are defined at fixed $L$ (the separation between two nuclei) with energies  $E_n(L)$, and certain levels cross each other at specific value of $L$.
The issue is the probability of the transition during such crossing of
two levels. Consider two levels with their energies given approximately by $E_1(L)\approx \sigma_1 L+C_1$ and $E_2(L)
\approx \sigma_2 L+C_2$ near the crossing point. When the two nuclei approach the crossing point adiabatically slowly $v=\dot L\rightarrow 0$, the
 electrons  always change from one state to the other
 selecting the  $lowest$  state at any $L$. If the two nuclei have fast relative motion then the transition between the two levels at crossing point is suppressed. More
quantitatively, Landau and Zener showed that the probability to remain
in the original state (i.e. no transition) is exponentially small at small velocity $v$
\begin{equation} \label{Landau_Zener}
P_{remain}=exp{\bigg [}  -\frac{2\pi |H_{12}|^2}{ v
|\sigma_1-\sigma_2|} {\bigg ]}
\end{equation}
where $H_{12}$ is the off-diagonal matrix element of
a two-level model Hamiltonian describing the transition between the two levels.

\section{Stable and Metastable Flux Tubes}

We now turn to the possible microscopic origin of the linear
rise in both potentials. Let's start with the ``dual superconductor'' model
 for QCD confinement in the vacuum, introduced by t'Hooft-Mandelstam \cite{'t Hooft-Mandelstam} and well supported by extensive studies in lattice QCD.
In this model, certain ``magnetically charged'' condensate (i.e.  a magnetic superconductor) occupies the vacuum
 and expels the electric flux between $\bar Q Q$ into a stable flux tube by forming magnetic {\em super-current} on tube surface, known as (dual)
 Meissner effect. Such flux tube naturally gives rise to a linear potential and the vacuum string tension is
thus identified with energy per unit length of such flux tube, mathematically
described by the well-known Abrikosov-Nielsen-Olesen (ANO) solution
\cite{Abrikosov} (for reviews and further references see e.g.
\cite{Ripka,Bali_review,Greensite,DiGiacomo}). What happens at finite T then?
With increasing T, the free energy tension decreases and eventually the linear part in
 $\bar Q Q$ free energy disappears at $T_c$, signaling the deconfinement transition. Since the flux tube and free energy tension
 is a direct consequence of the magnetic condensate, the decrease of $\sigma_F$ toward $T_c$ is naturally interpreted as the gradual ``melting'' of the magnetic condensate due to thermal excitations: similar phenomena is known for the usual superconductor in condensed matter systems.

Now, where does the linear part in internal energy (and the associated tension $\sigma_V$) come from? In particular, why does it persist even above $T_c$? The answer first proposed
in \cite{LS_monopole} relates it to the ``normal'' monopoles, as opposed to the BEC condensed monopoles existing only below $T_c$.  
 Such thermal monopoles can also expel the electric flux into (meta-stable) flux tube by forming a magnetic current (which may suffer from dissipation) on the tube surface: its dual phenomenon, i.e. magnetic flux tube formation in thermal electron plasma is
well know in classical (e.g. solar) plasma physics. Specific condition for the
persistence of the electric flux tube in a magnetic plasma was further developed in \cite{Liao:2007mj}, for infinitely long
flux tubes. There it has been found that ``normal'' monopoles are much less effective for this task as compared with ``super'' monopoles, but nevertheless able to mechanically stabilize the flux tube provided high enough density of these thermal monopoles. What has not been previously considered is the mechanism for dynamical formation of flux tube between a $\bar Q Q$ pair with finite separation.

\begin{figure}[h]
\begin{center}
\includegraphics[width=6.5cm]{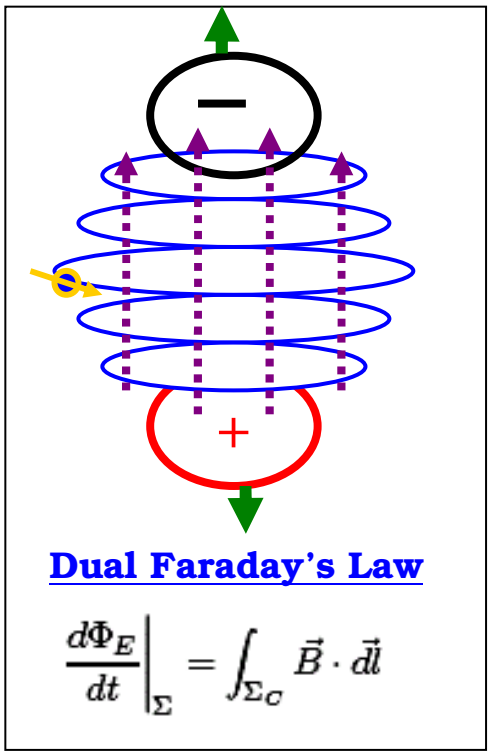}
\end{center}
\caption{\label{fig_dual}  Schematic demonstration of
magnetic solenoidal by Dural Faraday's law, see text. }
\end{figure}

Here we provide a dynamical explanation of why large energy,
growing approximately linearly with length, appears in a magnetic
plasma when a pair of two electric charges are separated with certain speed
$v$, see sketch of the setting in Fig.\ref{fig_dual}. The
answer lies in the Maxwell equations (with the presence of magnetic sources, see e.g.\cite{jackson}),
in particularly the $dual$ Faraday's law which relates the circulation
of the magnetic field $\int \vec B \vec dl$ over a closed contour with
the {\em change of electric flux} penetrating the enclosed area. As an electric charge moves
through the loop, rotating magnetic field in the magnetic medium leads
to solenoidal magnetic current (a ``magnetic coil''). In the
confined phase $T<T_c$ this current, after relaxation, becomes
the persistent super-current, remaining forever without loss:
thus  the free energy $F$  has a linear term for $T<T_c$.
 In a deconfined plasma phase $T>T_c$ this is impossible, thus
$\sigma_F=0$: the solenoidal ``magnetic coil''
 created in the fast process has only normal magnetic current, which is a meta-stable flux tube and eventually disappears
 due to dissipation. Yet it is still generated: thus $\sigma_V-\sigma_F$ is nonzero and there is splitting between free and internal energy.

Let us emphasize again the different roles of the
{\em super} and {\em normal} magnetic components. The former responds quantum
mechanically as a whole and does not  
 generate any entropy nor contribute to the splitting. the latter, however, has finite relaxation time and nonzero dissipation, ``feels'' the different time scales involved in the slow/fast processes, and therefore is responsible for entropy generation and the splitting between free and internal energy. In short, the $\sigma_F$ tells us
about the super component only, while the difference
$\sigma_V-\sigma_F$ tells us about the normal component.
One  arrives at the following picture for an  evolving
magnetic medium: with increasing $T$ the monopole ensemble
starts as a monopole condensate and continuously evaporates into a mixture
of both condensed and thermal monopoles;  at $T>T_c$ the condensate
melts entirely into a {\it normal} component of thermally excited monopoles. If so,
the thermal monopoles are expected to be most important in the temperature range $0.8-1.3\, T_c$ where the splitting is most significant.

We end this Section with discussions on a few important phenomenological implications
of the ``dual Faraday effect'' and the meta-stable flux tube. First, it means that magnetic monopoles may induce new mechanism of (electric) jet energy
loss particularly near $T_c$. In the jet quenching process, a very {\em fast} electric parton (quark or gluon) penetrates the bulk medium through various phases and thus may create behind it the above discussed ``magnetic coil'' in the near $T_c$ region where there are abundant thermal monopoles to be accelerated solenoidally by $B$ field due to the fast moving electric jet and thus take enormous amount of energy away from the jet. Such near-$T_c$ enhancement of jet quenching has been first suggested by us in \cite{Liao:2008dk} and found to be strongly favored by the azimuthal anisotropy data of jet quenching. Second, it also implies specific patterns of multi-particle correlations if such flux tubes can be created and protected by monopoles in heavy ion collisions, as elaborated first in \cite{Shuryak:2009cy}. One example is related to what happens to the flux tube created by a fast jet: clearly the monopoles forming the ``coil'' will subsequently
collide with the bulk thermal matter, with their energy being converted and distributed into the bigger volume: this may possibly be the beginning of ``conical flow'' process
suggested in \cite{CasalderreySolana:2004qm}. The other example concerns the narrow-azimuthal-angle long-range-rapidity correlations known as ``ridge'' which seems originating from certain local initial fluctuation seeds, but its narrowness in angle may be possibly preserved till the end of long bulk evolution only if certain mechanism like the flux tube by the thermal monopoles protects the initial seed from acoustic expansion (see detailed discussions in \cite{Shuryak:2009cy}). The existence of meta-stable flux tubes in the near $T_c$ plasma (and their associated large entropy) may also bear relevance to the observed cluster correlations \cite{Alver:2007wy}. While the existence and dynamical formation of such flux tubes are studied in \cite{Liao:2007mj} and here, another very important question (particularly for phenomenology) is its life time, i.e. the flux tube decay and the end products. This problem has recently been partially addressed in \cite{Faroughy:2010cd} where a relatively short life time is found in the classical treatment. On one hand from hydrodynamic modeling we know the near $T_c$ plasma has very small shear viscosity which indicates short mean free path and frequent scattering, while on the other hand for the magnetic currents to last long and hold the flux tube they better do not scatter too often: such a dilemma might be resolved if the thermal monopoles become really coherent over large distance at $T$ close to $T_c$ and the lattice study supports such coherence \cite{D'Alessandro:2010xg}. These questions will be studied further elsewhere.

\section{ Electric Field Solution in the Ellipsoidal Bag}

In this Section we will solve the Maxwell equation for electric field
induced by a pair of static charge-anti-charge separated by a distance
$L=2a$ along $\hat {\mathbf z}$ axis ($\pm Q_e$ sitting at $\mp a\hat z$), with a special ``tangent boundary condition'' (T.B.C.) on the boundary
surface $\Sigma_B$ , i.e.
\begin{eqnarray} \label{Laplace}
&& \vec{\bigtriangledown} ^2 \Phi(\br) = Q_e [\delta^3(\br - a \hat
{\mathbf z})-\delta^3(\br + a \hat  {\mathbf z})]  \\
&& \vec{\bigtriangledown} \Phi \cdot \hat{\mathbf
n}_{\Sigma_B} |_{\Sigma_B} = 0 \nonumber
\end{eqnarray}
The model itself  is a version of an old idea known as the Bag Model used
for light hadrons \cite{Chodos:1974je} at $T=0$, now generalized to
 give an approximate description of the electric field configuration between static $\bar{Q} Q$ in
 the chromo-magnetic medium at finite temperature.

A simplification we use is that the boundary $\Sigma_B$ is
approximated by a rotational ellipsoid with the two charges at its
focal points. This boundary shape can be specified by a single
parameter $\xi_B$, the ellipticity. Such boundary $\Sigma_B$ is
very conveniently parameterized in terms of the parabolic
coordinates system $(\xi,\eta,\phi)$, which we use: see Appendix A
for necessary formulae related to it. In Fig.\ref{fig_shape}
we show a few ellipsoidal shapes with parameters (from inside to
outside) $(L,\xi_B)$ to be
$(0.1,6.62)$,$(1,1.68)$,$(2,1.29)$,$(3,1.16)$ respectively, the
dashed lines indicate constant-$\eta$ curves (for $L=3$ case) with
(from top to bottom) $\eta=0.8,0.5,0.2,-0.2,-0.5,-0.8$, the
solid/empty circles indicate the positions of positive/negative
charges, and the arrows indicate the tangent electric fields on
the boundary.

\begin{figure}[h]
\begin{center}
\includegraphics[width=7.5cm]{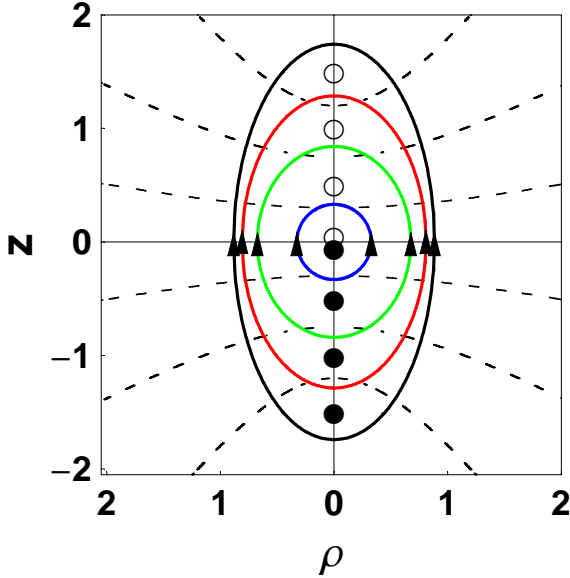}
\end{center}
\caption{\label{fig_shape} The ellipsoidal shapes we use for solving the electric field equations, see text for detailed explanations.\newline}
\end{figure}

We follow the standard method in classical electrostatics, see for example
\cite{jackson}. First, we rewrite (\ref{Laplace}) in
$(\xi,\eta,\phi)$ coordinates for solutions with axial symmetry, i.e. assuming $\Phi=\Phi(\xi,\eta)$ independent of angle $\phi$
\begin{eqnarray} \label{eq_elliptic_laplace}
&& \frac{\partial}{\partial \xi} {\bigg [} (\xi^2-1)\frac{\partial
\Phi}{\partial \xi} {\bigg ]} + \frac{\partial}{\partial \eta}
{\bigg [} (1-\eta^2)\frac{\partial \Phi}{\partial \eta} {\bigg ]}\nonumber \\
&&= \frac{Q_e \delta(\xi-1)}{\pi L} {\bigg [} \delta(\eta-1) -
\delta(\eta+1)  {\bigg ]} \nonumber \\
&&= \sum_{\nu=1,3,5,...}
\frac{Q_e \delta(\xi-1)}{\pi L}  (2\nu+1) P_\nu [\eta]
\end{eqnarray}
The last line in the above is an expansion of the $\eta$-dependence in terms of
Legendre functions $P_\nu [\eta]$ which in the interval $\eta \in
[-1,1]$ form a set of orthogonal and complete basis functions. Similarly, we do the expansion for the $\eta$-dependence of $\Phi$:
\begin{eqnarray}
\Phi_{\xi,\eta}=\sum_{\nu=1,3,5,...} \frac{Q_e f_\nu[\xi]}{\pi L}
(2\nu+1) P_\nu[\eta]
\end{eqnarray}
Then by simply comparing the coefficients of $P_\nu [\eta]$ on
 both sides of Eq.(\ref{eq_elliptic_laplace}) we obtain the
  equations for the functions $f_\nu[\xi]$ defined in $\xi \in (1,\infty)$:
\begin{eqnarray} \label{xi_eqn}
\frac{d}{d\xi} {\bigg [ }  (1-\xi^2) \frac{d f_\nu}{d \xi}  {\bigg
] } + \nu (\nu+1) f_\nu = - \delta (\xi-1)
\end{eqnarray}
while the boundary condition in Eq(\ref{Laplace}) now becomes
\begin{eqnarray}
f \, '[\xi=\xi_B]=0
\end{eqnarray}
with the parameter $\xi_B$ specifying the boundary surface $\Sigma_B$.
The solutions are given in terms of the
   Legendre functions of the first and second kinds:
\begin{eqnarray}\label{xi_solution}
&& f_\nu [\xi] = -k_\nu^B P_\nu [\xi] - Q_\nu [\xi] \\
&& k_\nu^B = -\frac{Q\, '_\nu [\xi_B]}{P\, '_\nu [\xi_B]} = -
\frac{\xi_B Q_\nu[\xi_B]-Q_{\nu-1}[\xi_B]}{\xi_B
P_\nu[\xi_B]-P_{\nu-1}[\xi_B]}  \nonumber
\end{eqnarray}
The full electrostatic potential is then given by
\begin{eqnarray}\label{the_potential}
&& \Phi{\big (} \vec{r}\, {\big |} L,\xi_B  {\big )} \nonumber \\
&& \quad =-\frac{Q_e}{4\pi L} \sum_{\nu=1,3,5,...} (8\nu+4) P_\nu[\eta]
{\big
(} k_\nu^B P_\nu [\xi] + Q_\nu [\xi]  {\big )} \nonumber \\
&& \quad = \frac{Q_e}{4\pi L} \frac{2}{\xi+\eta}+ \frac{(-Q_e)}{4\pi L}
\frac{2}{\xi-\eta} \nonumber \\
&&\quad \qquad - \frac{Q_e}{4\pi L} \sum_{\nu=1,3,5,...}
(8\nu+4) k_\nu^B P_\nu [\xi] P_\nu[\eta]
\end{eqnarray}
We've used the Neumann expansion of Legendre functions (see e.g.
\cite{Wang_Guo}) to write down the second equality: in there the
first two terms are nothing but the usual Coulomb potentials by
the $\pm Q_e$ charges, while the last summation term reflects the
nontrivial boundary contribution. At very large $\nu$ the summand
terms go asymptotically like $\nu \xi^{\nu} / \xi_B^{2\nu+2}$, so with $\xi$ satisfying $1<\xi\le \xi_B$, the summation is guaranteed to converge. The electric
field $\vec E = -\vec{\bigtriangledown} \Phi$ has been calculated using
(\ref{appen_gradient}) and the expression is quite lengthy which we skip showing here.

The volume occupied by the electric field (i.e. the ellipsoid
bulk within $\xi_B$) is given by
\begin{eqnarray} \label{xi_volume}
V_E(L , \xi_B) &=& \int_1^{\xi_B}d\xi \int_{-1}^1 d\eta
\int_0^{2\pi} d\phi \, H_\xi H_\eta H_\phi \nonumber \\
&=& \frac{\pi L^3 }{6}
\xi_B (\xi_B^2-1)
\end{eqnarray}
And the total electric field energy in this volume is given by
\begin{eqnarray} \label{xi_energy}
&&{\mathcal E}_{total} (L , \xi_B) \nonumber \\
&&\quad = \int_1^{\xi_B}d\xi
\int_{-1}^1 d\eta \int_0^{2\pi} d\phi \, H_\xi H_\eta H_\phi \,
\frac{\rho_e \times \Phi(\xi,\eta)}{2}\nonumber \\
&&\quad={\mathcal E}_{self} +
{\mathcal E}_{E}
\nonumber \\
&& {\mathcal E}_{self} = \frac{Q_e^2}{4\pi L}\frac{1}{(\xi+\eta)\to 0} + \frac{Q_e^2}{4\pi L}\frac{1}{(\xi-\eta)\to 0} \nonumber \\
&& {\mathcal E}_{E} = -\frac{Q_e^2}{4\pi L}  + \frac{Q_e^2}{4\pi
L} \sum_{\nu=1,3,5,...} (8\nu+4) k_\nu^B  \nonumber \\
&&\quad\,\,\, \equiv \frac{Q_e^2}{4\pi
L} \bar{\mathcal E}_E(\xi_B)
\end{eqnarray}
The ${\mathcal E}_{self}$ is the familiar self-interaction of the
two charges which we discard. The ``real'' interactional energy
${\mathcal E}_{E}$ consists (again) a Coulomb piece and a boundary
modification.

We conclude this section by one remark: so far the two key
variables $L$ and $\xi_B$ remain free parameters: they will be
related in the next section.

\section{The Free and Internal Energy of the Charge Pair}

With the solutions of electric field in the ellipsoidal bag (characterized by two parameters $L$ and $\xi_B$) from preceding Section, we now examine the dynamic formation of such bag when separating a pair of $\bar{Q} Q$ from zero to a finite distance $L$. The key point is that for a given $L$, the bag boundary $\xi_B$ shall be optimized so that the ``cost'' for creating such a configuration is minimized. Furthermore we study two settings: slow and fast
separation of the $\bar{Q} Q$ to a finite distance $L$, with the outcome being respectively the free and internal energy
associated with the pair. With slow separation, the free energy increase associated with the pair shall be minimized, and the dominant contributions to the free energy include both the electric field energy stored inside the bag and the energy needed to exclude the monopole condensate out of the bag volume (noting that for both there is no entropy associated and for thermal monopoles their contribution to free energy in the slow separation process largely cancels out between energy and entropy). With fast separation, the energy increase shall be minimized, and the dominant contributions to the energy include both the electric field energy stored inside the bag and the energy deposited to the thermal monopoles via the dual Faraday effect. We will calculate both processes in the rest of this Section and make connections with the lattice data.

\subsection{Free Energy from Slow Separation}

\begin{figure*}
\begin{center}
\includegraphics[height=6.cm]{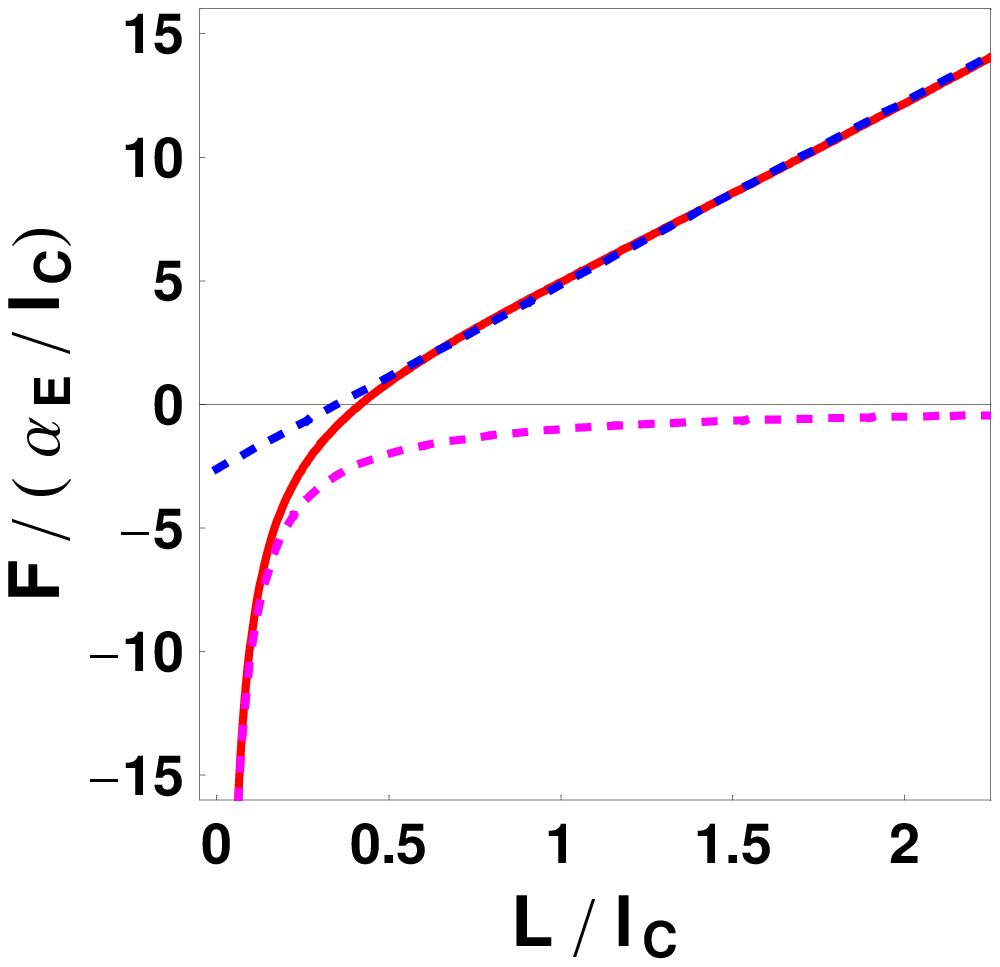}\hglue 10mm
\includegraphics[height=6.cm]{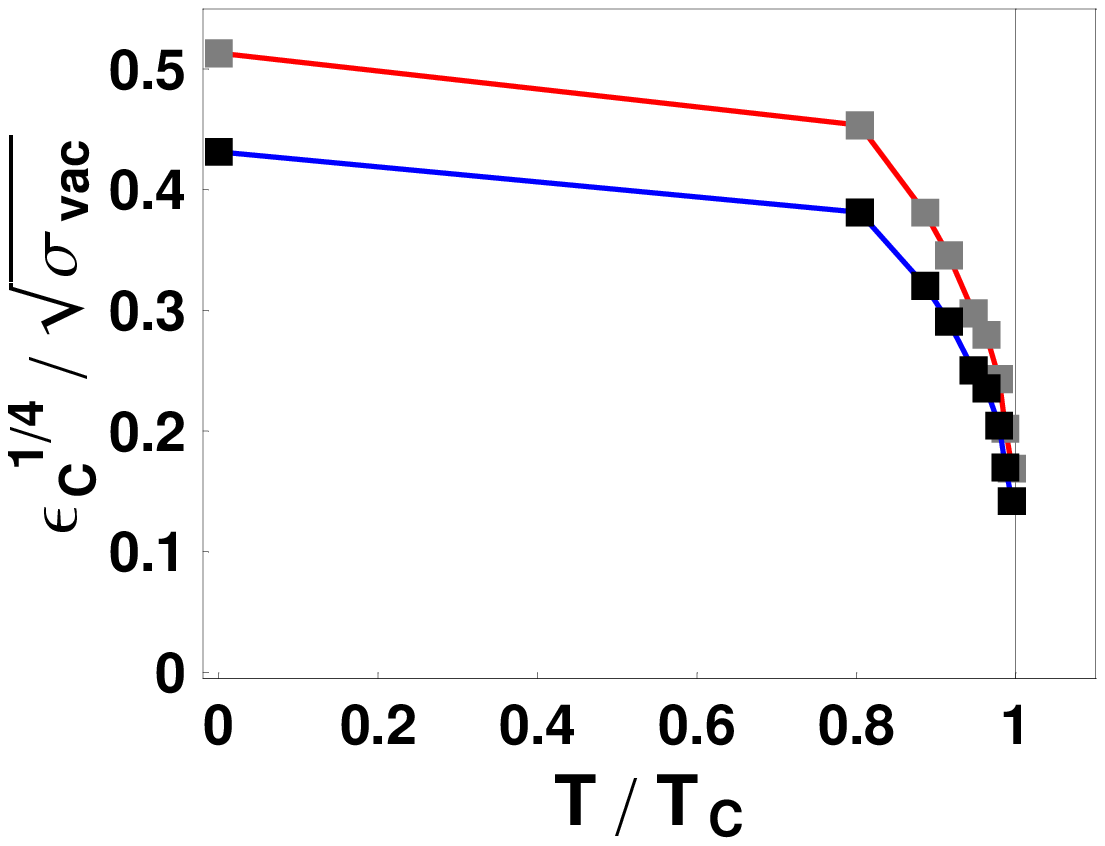}
\end{center}
\caption{\label{fig_free_energy} (a)(left) free energy $F$ (in
unit of $\alpha_E/l_C$) versus separation $L/l_C$; (b)(Right)
monopole condensate energy density $(\epsilon_C)^{1/4}$ in unit of
$\sqrt{\sigma_{vac}}$ with the two curves for $\alpha_E$ being
0.5(upper, red) and 1(lower, blue) in Eq.(\ref{tension_free}) respectively.}
\end{figure*}

As afore-discussed, when the $\bar{Q} Q$ pair is
separated in an adiabatically slow way, the {\it super} component
of the magnetic medium i.e. the monopole condensate will be
expelled entirely (in an idealized picture) out of the volume
$V_E$ occupied by electric field. Suppose the condensate has a
negative energy density $-{\epsilon}_C$ (thus a positive
``bag pressure''), then the overall change in free energy brought about by
separating the pair will be
\begin{eqnarray} \label{free_energy}
 \Delta F &&= {\mathcal E}_E(L,\xi_B) + {\epsilon}_C(T)\cdot V_E(L,\xi_B)
\end{eqnarray}
Now for given separation distance $L$ and bulk temperature $T$, we
determine the physical boundary of flux bag $\xi_B^{phy}$ by
minimizing the above $\Delta F$, i.e. the physical boundary
$\xi_B^{phy}(L,T)$ satisfies:
\begin{equation}
\frac{\partial \Delta F}{\partial \xi_B} {\bigg
|}_{\xi_B=\xi_B^{phy}} = 0
\end{equation}
Combining the above with Eq.(\ref{xi_volume},\ref{xi_energy}) we
then obtain
\begin{equation}
{\bigg [ }  \frac{1}{3\xi_B^2-1} \frac{d \bar{\mathcal
E}_E}{d\xi_B} {\bigg ]}  {\bigg |}_{\xi_B=\xi_B^{phy}} = - {\bigg
(} \frac{L}{l_C} {\bigg )} ^4
\end{equation}
where we have introduced a length scale
\begin{eqnarray}
l_C \equiv (6\, \alpha_E/ \pi \epsilon_C)^{1/4}
\end{eqnarray}
with $\alpha_E\equiv Q_e^2/4\pi$. This
equation could be solved easily by numerics. For each $L$ with the
above determined $\xi_B^{phy}$, we obtain via (\ref{free_energy})
the free energy associated with the pair as a function of
separation $L$, shown in Fig.\ref{fig_free_energy}(a). It turns
out to be Coulomb at short distance(see the magenta dashed curve)
plus linear at large distance(see the blue dashed line). The
occurrence of a linear part is due to the physical effect that for
large $L$ the medium pressure (with which the electric field has
to balance) limits the transverse size of flux bag (where the
field gets weak as $L$ increases) to saturate rather than grow
forever: thus the bag shape approaches a cylinder. Mathematically,
as $L\to \infty$ one finds $\xi_B^{phy}\to 1$ but $L \cdot
\sqrt{(\xi_B^{phy})^2-1} \to {\rm finite}$. In Fig.\ref{fig_shape}
the four bag shapes are at growing $L$ with $\xi_B$ determined as
in the above, which clearly
shows the shape becomes more and more cylindrical at large $L$. \\

By fitting the dimensionless slop of the linear part in
Fig.\ref{fig_free_energy}(a) we obtain the free energy string
tension $\sigma_F$:
\begin{equation} \label{tension_free}
\sqrt{\sigma_F}=2.32 \times \alpha_E^{1/4} \times \epsilon_C^{1/4}
\end{equation}
Inversely, since we know $\sigma_F(T)$ from lattice as shown in
Fig.\ref{fig_tension}, from the above formula we can infer the
$T$-dependence of the monopole condensate energy density $\epsilon_C$: see
Fig.\ref{fig_free_energy}(b). The two curves are for $\alpha_E$
being 0.5(upper) and 1(lower) respectively. In both cases,
$\epsilon_C$ decreases with $T$ and drops abruptly
close to $T_c$. The interpretation is natural: toward
$T_c$ the monopole condensate becomes less and less due to
increasing thermal excitations and eventually dies out around
$T_c$.

A connection can be made between our result (\ref{tension_free})
and the dual superconductor model (also known as Abelian Higgs
model) of vacuum confinement \cite{Ripka}. In that model, a
quadratic Higgs potential leads to a Higgs condensate (the
prototype of postulated monopole condensate) $\phi_0$ (with
dimension of mass). By solving ANO flux tube a string tension is
obtained in the form $\sqrt{\sigma}=c_1 \phi_0$ with the
coefficient determined by gauge and Higgs coupling constants
$\lambda$ and $g$. On the other hand the Higgs potential implies that the
condensate has a negative energy density $-\epsilon_C=-\lambda
\phi_0^4/2$, thus one arrives at a similar relation between string tension and condensate energy density: $\sqrt{\sigma}=c_2 \epsilon_C^{1/4}$ in that model with the coefficient to be determined numerically for given
coupling parameters, see e.g. \cite{Baker:1998jw}. While
that model works primarily at $T=0$, our model for $\sigma_F$ extends
to finite T.

\begin{figure*}
\begin{center}
\includegraphics[height=6.cm]{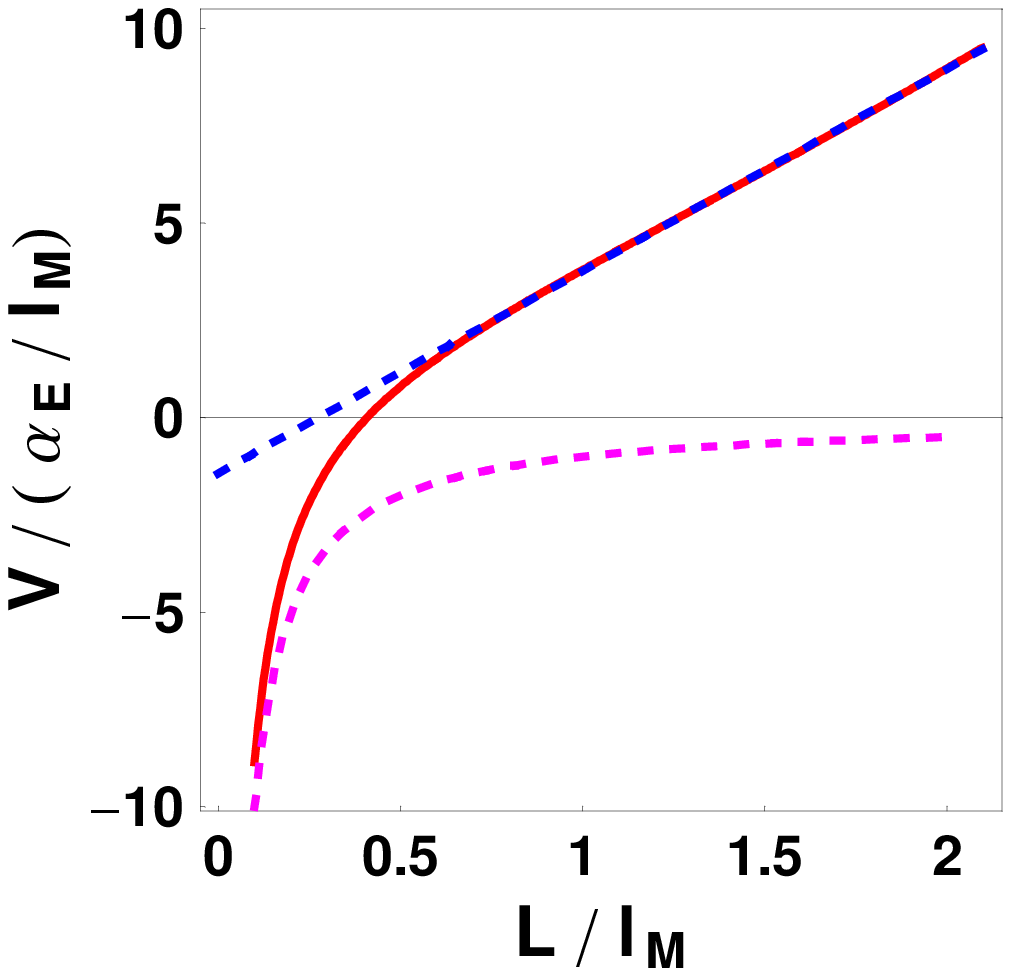}\hglue 10mm
\includegraphics[height=6.cm]{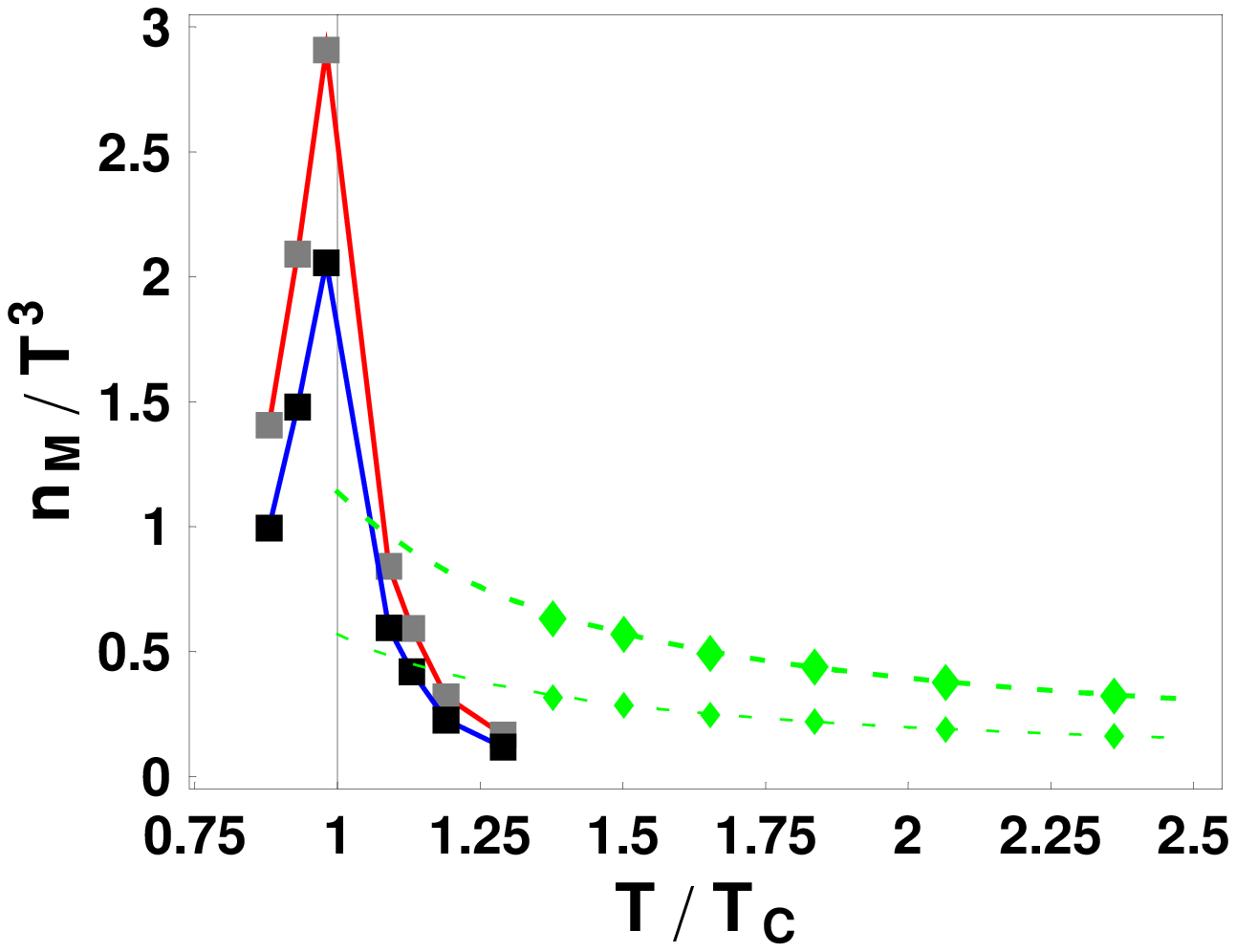}
\end{center}
\caption{\label{fig_potential_energy} (a)(left) internal energy
$V$ (in unit of $\alpha_E/l_M$) versus separation $L/l_M$;
(b)(Right) thermal monopole density $n_M/T^3$ with the two curves connecting box symbols for $\alpha_E$ being 0.5(upper, red) and 1(lower, blue) in Eq.(\ref{tension_fast})
respectively, while the two green curves connecting diamond symbols showing $SU(2)$(lower) lattice data and their $SU(3)$ extrapolation(upper) for
$T>1.3T_c$ from \cite{lattice_mag_density} (see text for more details).}
\end{figure*}

\subsection{Internal Energy from Fast Separation}

Now we study the case of separating the two charges to a finite
distance $L$ within a time much smaller than the relaxation time
of the surrounding thermal bath. In
particular we focus on the region about $0.8-1.3T_c$, in which the {\it normal} component of thermal monopoles becomes substantial and dominant while the {\it super} component becomes less and less.

During such fast process, each monopole originally in the volume
to be occupied by the electric filed (i.e. the ellipsoidal bag)
will get a "kick" due to the dual Faraday effect (see Fig.\ref{fig_dual}) but have no time
to release this energy into the surrounding medium. Suppose the
positive charge is moved along $\hat z$-axis from $z=C$ to $z=C+\delta z$ in $\delta t$ (and correspondingly the negative one from $z=-C$ to $z=-(C+\delta z)$), then the electric flux penetrating the plane $z=C$ changes from 0 to $Q_e$,
thus generating a magnetic dynamical voltage $Q_e/\delta t$. For a
monopole at a transverse distance $\rho$ from $\hat z$ axis, the force is
$Q_m(Q_e/\delta t)/(2\pi \rho)$, thus it gets the "kick" and
obtains a momentum $\delta p = Q_m Q_e /( 2\pi \rho)$, forming
strong {\it non-thermal} and {\it non-super} magnetic currents. For a bag $(L,\xi_B)$
formed after separation, the total kinetic energy passed to the
monopoles in the flux bag is obtained by integration over the bag volume (with
$D\equiv Q_m Q_e / 4\pi=1$):
\begin{eqnarray}
\Delta K_M &=& \int_1^{\xi_B}d\xi \int_{-1}^1 d\eta \int_0^{2\pi}
d\phi \, H_\xi H_\eta H_\phi \nonumber \\
&&\qquad \times \frac{4 n_M \,
D}{L\sqrt{(\xi^2-1)(1-\eta^2)}} \nonumber \\
&=& \frac{\pi^2 D n_M}{2} L^2 \xi_B
(\xi_B^2-1)^{1/2}
\end{eqnarray}
We emphasize in the above only the monopole density $n_M(T)$
enters as a property of the medium depending on $T$, while other
equilibrium properties of the medium shall not be "felt" in such fast process.

Now the total energy change during such process includes the electric field energy in the bag volume, the energy for expelling the monopole condensate (if there is any) out of the volume, and the kinetic energy delivered to the normal monopoles, which are summed to be $\Delta E ={\mathcal
E}_E(L,\xi_B)+{\epsilon}_C V_E+ \Delta K_M (L,\xi_B)$. The last
new term due to the thermal monopoles will then partly convert into entropy  after they interact with the medium particles at large for a time longer than the relaxation, ultimately causing the splitting between free/internal energy.
Since the condensate term ${\epsilon}_C V_E$ becomes very small close to
$T_c$ (as we showed in previous subsection) and vanishes above
$T_c$, we neglect it here for simplicity, i.e. $\Delta E \approx
{\mathcal E}_E(L,\xi_B)+ \Delta K_M (L,\xi_B)$. To obtain the
physical value $\xi_B^{phy}$, we need to minimize $\Delta E$
according to $\xi_B$, which leads to
\begin{equation}
\frac{\partial \Delta E}{\partial \xi_B} {\bigg
|}_{\xi_B=\xi_B^{phy}} = 0
\end{equation}
This can be written as
\begin{equation}
{\bigg [ }  \frac{\sqrt{\xi_B^2-1}}{2\xi_B^2-1} \frac{d
\bar{\mathcal E}_E}{d\xi_B} {\bigg ]}  {\bigg
|}_{\xi_B=\xi_B^{phy}} = - {\bigg (} \frac{L}{l_M} {\bigg )} ^3
\end{equation}
Here we introduced a different length scale $l_M \equiv (2\,
\alpha_E/ \pi^2 D n_M)^{1/3}$. Not surprisingly we find the
internal energy, shown in Fig.\ref{fig_potential_energy}(a), to
be a Coulomb at short distance(see the magenta dashed curve) plus
linear at large distance(see the blue dashed line).

Now the string tension in the internal energy is given by the following formula:
\begin{equation} \label{tension_fast}
\sqrt{\sigma_V}=3.88 \times \alpha_E^{1/6} \times n_M^{1/3}
\end{equation}
Since we know $\sigma_V$ from lattice data in the $0.8-1.3T_c$
region, by the above formula we can convert $\sigma V$ into
thermal monopole density $n_M(T)$ in the same region: see the two
curves(connecting box symbols) for $\alpha_E$ being 0.5(upper, red) and 1(lower, blue)
respectively in Fig.\ref{fig_potential_energy}(b). We also show an independent information on the thermal monopole density above $1.3T_c$ from lattice study in \cite{lattice_mag_density} in Fig.\ref{fig_potential_energy}(b) as green curves(connecting diamond symbols): the lower one is the original data for $SU(2)$ in \cite{lattice_mag_density}, while the upper one is an extrapolation to $SU(3)$ by the simple $N_c-1$ scaling for monopole species (i.e. twice more monopoles in $SU(3)$ than in $SU(2)$), with both curves extended toward $T_c$ according to the fitting formula $n_m/T^3=0.48/log(2.48\cdot T/T_c)^{1.89}$ (and twice in the upper one for $SU(3)$) in \cite{lattice_mag_density}. The comparison shows reasonably good agreement between our estimates for the thermal monopole density from string tension and the measured density by directly
identifying the thermal monopoles on the lattice.

A few comments are in order: (i) for $0.8-1\,T_c$ the density
quickly grows toward $T_c$ while at the same time results from previous subsection show rapid dropping of condensate density in the same region, which strongly indicates the scenario that close to $T_c$ monopole condensate
is continuously and substantially getting excited into thermal monopoles; (ii) around
$1.3T_c$ we see our results connect well to the higher $T$ lattice
data with reasonable values of coupling $\alpha_E$; (iii) cooling down to $T_c$ we find the monopole density $n_M/T^3$ quickly rising almost by an order of magnitude; (iv) the
strongly increasing density also suggests rapid increase of
magnetic screening toward $T_c$, which is in agreement with lattice
results \cite{Nakamura}.

A particularly interesting feature is that the (normalized) density $n_M/T^3$
increases roughly by {\it one order of magnitude} from $1.3T_c$ down to
$T_c$, with the number much larger than even a Stefan-Boltzman gas. This indicates that near $T_c$ the monopoles should be very light, and furthermore their interactions should make it beneficial in energy to have a large number of monopole-anti-monopole pairs. The monopoles are so dense and light that they become the dominant component in the near-$T_c$ plasma
and presumably become quantum coherent, and eventually reach the condensation point at $T_c$ (see most recent lattice results in \cite{D'Alessandro:2010xg} showing evidences for such a scenario). It has been suggested that these thermal monopoles near $T_c$
seem to form a densely packed liquid \cite{Liao:2008jg}\cite{lattice_mag_density}.

\section{Summary}

In this paper, we have argued that the free energy $F(r,T)$  and internal energy $V(r,T)$
can be probed by slow and fast separation of the $\bar Q Q$ pair, respectively.
Furthermore we have identified the linear part in both potentials with flux tube formation between the pair: for free energy as probed by slow separation, there is stable flux
tube protected by magnetic super-current due to condensed monopoles which has no dissipation and exists below $T_c$; for internal energy as probed by fast separation, there is meta-stable flux tube protected by magnetic normal current due to thermal monopoles which are very dense in the region $0.8-1.3\, T_c$ and generate large entropy (the splitting between free and internal energy) via dissipation on longer time scale. Based on these ideas we have solved analytically the elliptic bags and provided expressions for the potentials at all separations, which happen to describe the data very well.

The main outcome from our study of the static $\bar Q Q$ potentials
is the particular relations we suggest between the free energy $F(r,T)$  and internal energy $V(r,T)$ measured on the lattice and
the densities of the condensed and ``normal" monopoles: see Eqs.(\ref{tension_free}) and (\ref{tension_fast}). Since those densities can be directly obtained from the lattice configurations,
one may check if these relations are correct or not. Such further tests of the
``magnetic scenario''\cite{LS_monopole,lattice_mag} for the near $T_c$
QCD plasma are rather straightforward and should be performed.

\bigskip

\section*{Acknowledgements}

The work of JL was partially supported by the Director, Office of Energy Research,
Office of High Energy and Nuclear Physics, Divisions of Nuclear
Physics, of the U.S. Department of Energy under Contract No.
DE-AC02-05CH11231. The work of ES was supported
in parts by the US-DOE grant DE-FG-88ER40388.

\section*{Appendix A}
\renewcommand{\theequation}{A\arabic{equation}}
\setcounter{equation}0 In this Appendix we briefly list the
parabolic coordinates formulae needed for the calculation in
Sec.2.

The coordinates we use are $(\xi,\eta,\phi)$ with two focal points
at $\pm a \hat{\bf z}$, which are related to cylindrical
coordinates $(\rho,\phi,z)$ by
\begin{equation} \label{appen_coor}
\rho= a\sqrt{(\xi^2-1)(1-\eta^2)} \quad , \quad \phi=\phi \quad ,
\quad z=a\xi \eta
\end{equation}
The variables are defined in the following domains: $\xi \in (1,\infty)$, $\eta \in
[-1,1]$, $\phi \in [0,2\pi)$. Writing $ds^2=H_\xi^2
d\xi^2+H_\eta^2 d\eta^2+H_\phi^2 d\phi^2$, we have
\begin{eqnarray} \label{appen_metric}
&& H_\xi= a \frac{\sqrt{\xi^2-\eta^2}}{\sqrt{\xi^2-1}} \quad , \quad
H_\eta= a \frac{\sqrt{\xi^2-\eta^2}}{\sqrt{1-\eta^2}} ,\nonumber \\
&& H_\phi=a\sqrt{(\xi^2-1)(1-\eta^2)}
\end{eqnarray}
The Laplacian is given by
\begin{eqnarray} \label{appen_laplacian}
{\vec \bigtriangledown}^2=\frac{1}{a^2 (\xi^2-\eta^2)}&& {\bigg \{}
\frac{\partial}{\partial \xi}{\bigg [}
(\xi^2-1)\frac{\partial}{\partial \xi} {\bigg ]}\nonumber \\
&& +
\frac{\partial}{\partial \eta}{\bigg [}
(1-\eta^2)\frac{\partial}{\partial \eta} {\bigg ]}\nonumber \\
&&+ {\bigg [ }
\frac{1}{\xi^2-1}+\frac{1}{1-\eta^2}  {\bigg
]}\frac{\partial^2}{\partial \phi^2} {\bigg \}}\quad
\end{eqnarray}
Finally the gradient is given by
\begin{equation} \label{appen_gradient}
{\vec \bigtriangledown}= \hat{\bf \xi} \frac{\partial}{H_\xi
\partial \xi} + \hat{\bf \eta} \frac{\partial}{H_\eta \partial \eta}
+ \hat{\bf \phi} \frac{\partial}{H_\phi
\partial \phi}
\end{equation}
For more details one could consult books such as \cite{Wang_Guo}.

%
% BibTeX users please use
% \bibliographystyle{}
% \bibliography{}

\begin{thebibliography}{}


\bibitem{Bali_review}
  G.~S.~Bali,
  %``QCD forces and heavy quark bound states,''
  Phys.\ Rept.\  {\bf 343}, 1 (2001)
  [arXiv:hep-ph/0001312].


%\cite{Greensite:2003bk}
\bibitem{Greensite}
  J.~Greensite,
  %``The confinement problem in lattice gauge theory,''
  Prog.\ Part.\ Nucl.\ Phys.\  {\bf 51}, 1 (2003)
  [arXiv:hep-lat/0301023];
    R.~Alkofer and J.~Greensite,
  %``Quark Confinement: The Hard Problem of Hadron Physics,''
  J.\ Phys.\ G {\bf 34}, S3 (2007)
  [arXiv:hep-ph/0610365].



\bibitem{Kaczmarek_Zantow}
    O.~Kaczmarek and F.~Zantow,
  %``Static quark anti-quark interactions in zero and finite temperature  QCD.
  %I: Heavy quark free energies, running coupling and quarkonium  binding,''
  Phys.\ Rev.\  D {\bf 71}, 114510 (2005)
  [arXiv:hep-lat/0503017];
  %``Static quark anti-quark interactions at zero and finite temperature  QCD.
  %II: Quark anti-quark internal energy and entropy,''
  arXiv:hep-lat/0506019.

\bibitem{Kaczmarek_pure_gauge}
  O.~Kaczmarek, F.~Karsch, E.~Laermann and M.~Lutgemeier,
  %``Heavy quark potentials in quenched QCD at high temperature,''
  Phys.\ Rev.\  D {\bf 62}, 034021 (2000)
  [arXiv:hep-lat/9908010];
   O.~Kaczmarek, F.~Karsch, P.~Petreczky and F.~Zantow,
  %``Heavy quark anti-quark free energy and the renormalized Polyakov loop,''
  Phys.\ Lett.\  B {\bf 543}, 41 (2002)
  [arXiv:hep-lat/0207002];
   O.~Kaczmarek, F.~Karsch, P.~Petreczky and F.~Zantow,
  %``Heavy quark free energies, potentials and the renormalized Polyakov
  %loop,''
  Nucl.\ Phys.\ Proc.\ Suppl.\  {\bf 129}, 560 (2004)
  [arXiv:hep-lat/0309121].


\bibitem{Petreczky_3f}
  P.~Petreczky and K.~Petrov,
  %``Free energy of a static quark anti-quark pair and the renormalized
  %Polyakov loop in three flavor QCD,''
  Phys.\ Rev.\  D {\bf 70}, 054503 (2004)
  [arXiv:hep-lat/0405009].


\bibitem{Kaczmarek_proceedings}
  O.~Kaczmarek and F.~Zantow,
  %``Static quark anti-quark free and internal energy in 2-flavor QCD and bound
  %states in the QGP,''
  PoS {\bf LAT2005}, 192 (2006)
  [arXiv:hep-lat/0510094].

\bibitem{Young:2008he}
  C.~Young and E.~Shuryak,
  %``Charmonium in strongly coupled quark-gluon plasma,''
  Phys.\ Rev.\  C {\bf 79}, 034907 (2009)
  [arXiv:0803.2866 [nucl-th]].
  %%CITATION = PHRVA,C79,034907;%%
%\cite{Young:2009tj}
%\bibitem{Young:2009tj}
  C.~Young and E.~Shuryak,
  %``Recombinant Charmonium in strongly coupled Quark-Gluon Plasma,''
  arXiv:0911.3080 [nucl-th].
  %%CITATION = ARXIV:0911.3080;%%

\bibitem{Zhao:2007hh}
  X.~Zhao and R.~Rapp,
  %``Transverse Momentum Spectra of J/\psi in Heavy-Ion Collisions,''
  Phys.\ Lett.\  B {\bf 664}, 253 (2008)
  [arXiv:0712.2407 [hep-ph]].

%\cite{Mocsy:2007yj}
\bibitem{Mocsy:2007yj}
  A.~Mocsy and P.~Petreczky,
  %``Can quarkonia survive deconfinement ?,''
  Phys.\ Rev.\  D {\bf 77}, 014501 (2008)
  [arXiv:0705.2559 [hep-ph]].


\bibitem{bound_states}
 E.~V.~Shuryak and I.~Zahed,
  %``Rethinking the properties of the quark gluon plasma at T approx. T(c),''
  Phys.\ Rev.\  C {\bf 70}, 021901 (2004)
  [arXiv:hep-ph/0307267];
   %``Towards a theory of binary bound states in the quark gluon plasma,''
  Phys.\ Rev.\  D {\bf 70}, 054507 (2004)
  [arXiv:hep-ph/0403127].

  \bibitem{LS_bound}
    J.~Liao and E.~V.~Shuryak,
  %``Polymer chains and baryons in a strongly coupled quark-gluon plasma,''
  Nucl.\ Phys.\  A {\bf 775}, 224 (2006)
  [arXiv:hep-ph/0508035].

\bibitem{susceptibilities}
  C.~R.~Allton {\it et al.},
  %``Thermodynamics of two flavor QCD to sixth order in quark chemical
  %potential,''
  Phys.\ Rev.\  D {\bf 71}, 054508 (2005)
  [arXiv:hep-lat/0501030];   R.~V.~Gavai and S.~Gupta,
  %``Simple patterns for non-linear susceptibilities near T(c),''
  Phys.\ Rev.\  D {\bf 72}, 054006 (2005)
  [arXiv:hep-lat/0507023].



\bibitem{LS_susceptibilities}
  J.~Liao and E.~V.~Shuryak,
  %``What do lattice baryonic susceptibilities tell us about quarks,  diquarks
  %and baryons at T > T(c)?,''
  Phys.\ Rev.\  D {\bf 73}, 014509 (2006)
  [arXiv:hep-ph/0510110].




\bibitem{Boris_Edward_Ismail}
B.~A.~Gelman, E.~V.~Shuryak and I.~Zahed,
  %``Classical Strongly Coupled QGP I: The Model and Molecular Dynamics
  %Simulations,''
Phys. Rev. C {\bf 74}, 044908 (2006)[nucl-th/0601029];
  %%CITATION = NUCL-TH 0601029;%%
Phys. Rev. C {\bf 74}, 044909 (2006)[nucl-th/0605046].
  %%CITATION = NUCL-TH 0605046;%%

\bibitem{LS_monopole}
  J.~Liao and E.~Shuryak,
  %``Strongly coupled plasma with electric and magnetic charges,''
  Phys.\ Rev.\  C {\bf 75}, 054907 (2007)
  [arXiv:hep-ph/0611131].
  %%CITATION = PHRVA,C75,054907;%%

  \bibitem{Ratti:2008jz}
  C.~Ratti and E.~Shuryak,
  %``The role of monopoles in a Gluon Plasma,''
  arXiv:0811.4174 [hep-ph].

\bibitem{sQGP_review}
M.~Gyulassy and L.~McLerran, Nucl. Phys. A {\bf 750}, 30 (2005);
E.~V.~Shuryak, Prog. Part. Nucl. Phys.{\bf 53}, 273 (2004)
[hep-ph/0312227]; Nucl.\ Phys.\ A {\bf 750}, 64 (2005).
   %%CITATION = HEP-PH 0312227;%%
  %``What RHIC experiments and theory tell us about properties of quark-gluon
  %plasma?,''

  \bibitem{sQGP_ES}
    E.V.Shuryak,
 %``Strongly coupled quark-gluon plasma: The status report,''
  arXiv:hep-ph/0608177;
  %``Emerging theory of strongly coupled quark-gluon plasma,''
  arXiv:hep-ph/0703208.

\bibitem{Antonov:2006wz}
  D.~Antonov, S.~Domdey and H.~J.~Pirner,
  %``A Heavy Quark-Antiquark Pair in Hot QCD,''
  Nucl.\ Phys.\  A {\bf 789}, 357 (2007)
  [arXiv:hep-ph/0612256].

\bibitem{Megias}
 E.~Megias, E.~Ruiz Arriola and L.~L.~Salcedo,
  %``The quark-antiquark potential at finite temperature and the dimension   two
  %gluon condensate,''
  Phys.\ Rev.\  D {\bf 75}, 105019 (2007)
  [arXiv:hep-ph/0702055].

%\cite{Beraudo:2007ky}
\bibitem{Beraudo:2007ky}
  A.~Beraudo, J.~P.~Blaizot and C.~Ratti,
  %``Real and imaginary-time $Q\bar{Q}$ correlators in a thermal medium,''
  arXiv:0712.4394 [nucl-th].
  %%CITATION = ARXIV:0712.4394;%%



\bibitem{Liao:2008jg}
  J.~Liao and E.~Shuryak,
%  ``Magnetic component of quark-gluon plasma is also a liquid!''
 Phys.\ Rev.\ Lett.\  {\bf 101}, 162302 (2008)
  [arXiv:0804.0255 [hep-ph]].
  %%CITATION = PHRVA,C75,054907;%%


\bibitem{lattice_mag}
  M.~N.~Chernodub and V.~I.~Zakharov,
  %``Magnetic component of Yang-Mills plasma,''
  Phys.\ Rev.\ Lett.\  {\bf 98}, 082002 (2007)
  [arXiv:hep-ph/0611228];   %``Magnetic strings as part of Yang-Mills plasma,''
  arXiv:hep-ph/0702245.



\bibitem{lattice_mag_eos}
  M.~N.~Chernodub, K.~Ishiguro, A.~Nakamura, T.~Sekido, T.~Suzuki and V.~I.~Zakharov,
  %``Topological defects and equation of state of gluon plasma,''
  PoS {\bf LAT2007}, 174 (2007)
  [arXiv:0710.2547 [hep-lat]].

\bibitem{lattice_mag_density}
  A.~D'Alessandro and M.~D'Elia,
  %``Magnetic monopoles in the high temperature phase of Yang-Mills theories,''
  arXiv:0711.1266 [hep-lat].


%\cite{Shuryak:2008pz}
\bibitem{Shuryak:2008pz}
  E.~Shuryak,
  %``Quark-Gluon Plasma - New Frontiers,''
  arXiv:0804.1373 [hep-ph].
  %%CITATION = ARXIV:0804.1373;%%
%\bibitem{Shuryak:2008eq}
  E.~Shuryak,
  %``Physics of Strongly coupled Quark-Gluon Plasma,''
  Prog.\ Part.\ Nucl.\ Phys.\  {\bf 62}, 48 (2009)
  [arXiv:0807.3033 [hep-ph]].
%\bibitem{Shuryak:2009zz}
  E.~Shuryak,
  %``Four lectures on strongly coupled Quark Gluon Plasma,''
  Nucl.\ Phys.\ Proc.\ Suppl.\  {\bf 195}, 111 (2009).
  %%CITATION = NUPHZ,195,111;%%


%\cite{Liao:2008wb}
\bibitem{Liao:2008wb}
J.~Liao and E.~Shuryak,
  %``Magnetic Component of Quark-Gluon Plasma,''
  J.\ Phys.\ G {\bf 35}, 104058 (2008)
  [arXiv:0804.3102 [hep-ph]].
%\cite{Liao:2008pu}
%\bibitem{Liao:2008pu}
  J.~Liao and E.~Shuryak,
  %``Magnetic Scenario for the QCD Fluid at RHIC,''
  arXiv:0809.2419 [hep-ph].
  %%CITATION = ARXIV:0809.2419;%%


%\cite{Liao:2007mj}
\bibitem{Liao:2007mj}
  J.~Liao and E.~Shuryak,
  %``Electric Flux Tube in Magnetic Plasma,''
  Phys.\ Rev.\  C {\bf 77}, 064905 (2008)
  [arXiv:0706.4465 [hep-ph]].
  %%CITATION = ARXIV:0706.4465;%%


\bibitem{Landau_QM}
L.~D.~Landau and E.~M.~Lifshitz, {\it ``Quantum Mechanics''}, 3rd
ed., Butterworth-Heinemann, 1981.

\bibitem{Landau_Zener}
L.~D.~Landau, Physics of the Soviet Union 2: 46¨C51 (1932).
C.~Zener, Proceedings of the Royal Society of London, Series A 137 (6): 696¨C702 (1932).
For a concise introduction see e.g.: C. Wittig, %``The Landau-Zener Formula''
J. Phys. Chem. B {\bf 109}, 8428 (2005).



\bibitem{'t Hooft-Mandelstam}
S.~Mandelstam,
  %``Vortices And Quark Confinement In Nonabelian Gauge Theories,''
  Phys.\ Rept.\  {\bf 23}, 245 (1976);  %%CITATION = PRPLC,23,245;%%
G.~'t Hooft,
   ``Topology Of The Gauge Condition And New Confinement Phases In Nonabelian
  %Gauge Theories,''
  Nucl.\ Phys.\ B {\bf 190}, 455 (1981).
  %%CITATION = NUPHA,B190,455;%%

\bibitem{Abrikosov} A.A. Abrikosov, Sov. Phys. JETP {\bf 32}, 1442
(1957); H.B. Nielsen and P. Olesen, Nucl. Phys. B {\bf 61}, 45
(1973).

\bibitem{Ripka}
  G.~Ripka, ``Dual superconductor models of color confinement'', Lecture notes in physics, Vol.
  639, Springer Berlin / Heidelberg 2004; arXiv:hep-ph/0310102.

%\cite{DiGiacomo:1998pm}
\bibitem{DiGiacomo}
  A.~Di Giacomo,
  %``Monopole condensation and colour confinement,''
  Prog.\ Theor.\ Phys.\ Suppl.\  {\bf 131}, 161 (1998)
  [arXiv:hep-lat/9802008];
    A.~Di Giacomo,
  %``Confinement of color: A review,''
  arXiv:hep-lat/0310023.
  %%CITATION = HEP-LAT/0310023;%%


\bibitem{Liao:2008dk}
  J.~Liao and E.~Shuryak,
  %``Where are jets quenched in heavy-ion collisions?,''
     Phys.\ Rev.\ Lett.\  {\bf 102}, 202302 (2009)
  [arXiv:0810.4116 [nucl-th]].
  %%CITATION = ARXIV:0810.4116;%%


%\cite{Shuryak:2009cy}
\bibitem{Shuryak:2009cy}
  E.~Shuryak,
  %``The Fate of the Initial State Perturbations in Heavy Ion Collisions,''
  Phys.\ Rev.\  C {\bf 80}, 054908 (2009)
  [Erratum-ibid.\  C {\bf 80}, 069902 (2009)]
  [arXiv:0903.3734 [nucl-th]].
  %%CITATION = PHRVA,C80,054908;%%

%\cite{CasalderreySolana:2004qm}
\bibitem{CasalderreySolana:2004qm}
  J.~Casalderrey-Solana, E.~V.~Shuryak and D.~Teaney,
  %``Conical flow induced by quenched QCD jets,''
  J.\ Phys.\ Conf.\ Ser.\  {\bf 27}, 22 (2005)
  [Nucl.\ Phys.\  A {\bf 774}, 577 (2006)]
  [arXiv:hep-ph/0411315].
  %%CITATION = NUPHA,A774,577;%%



%\cite{Alver:2007wy}
\bibitem{Alver:2007wy}
  B.~Alver {\it et al.}  [PHOBOS Collaboration],
  %``Cluster properties from two-particle angular correlations in p + p
  %collisions at s**(1/2) = 200-GeV and 410-GeV,''
  Phys.\ Rev.\  C {\bf 75}, 054913 (2007)
  [arXiv:0704.0966 [nucl-ex]].
  %%CITATION = PHRVA,C75,054913;%%
 B.~Alver {\it et al.}  [PHOBOS Collaboration],
  %``System size dependence of cluster properties from two-particle angular
  %correlations in Cu+Cu and Au+Au collisions at $\sqrt{s_{_{NN}}}$ = 200 GeV,''
  Phys.\ Rev.\  C {\bf 81}, 024904 (2010)
  [arXiv:0812.1172 [nucl-ex]].



%\cite{Faroughy:2010cd}
\bibitem{Faroughy:2010cd}
  C.~Faroughy and E.~Shuryak,
  %``The Lifetime of the Electric Flux Tubes near the QCD Phase Transition,''
  arXiv:1004.2890 [hep-ph].
  %%CITATION = ARXIV:1004.2890;%%

\bibitem{D'Alessandro:2010xg}
  A.~D'Alessandro, M.~D'Elia and E.~V.~Shuryak,
  %``Thermal Monopole Condensation and Confinement in finite temperature
  %Yang-Mills Theories,''
  Phys.\ Rev.\  D {\bf 81}, 094501 (2010)
  [arXiv:1002.4161 [hep-lat]].

%\cite{Chodos:1974je}
\bibitem{Chodos:1974je}
  A.~Chodos, R.~L.~Jaffe, K.~Johnson, C.~B.~Thorn and V.~F.~Weisskopf,
  %``A New Extended Model Of Hadrons,''
  Phys.\ Rev.\  D {\bf 9}, 3471 (1974).
  %%CITATION = PHRVA,D9,3471;%%


    \bibitem{jackson}
    J. D. Jackson, {\it Classical Electrodynamics} (3rd edition),
    John Wiley \& Sons, Inc. (1999).


\bibitem{Baker:1998jw}
  M.~Baker, N.~Brambilla, H.~G.~Dosch and A.~Vairo,
  %``Field strength correlators and dual effective dynamics in {QCD},''
  Phys.\ Rev.\  D {\bf 58}, 034010 (1998)
  [arXiv:hep-ph/9802273].
  %%CITATION = PHRVA,D58,034010;%%



%\cite{Nakamura:2003pu}
\bibitem{Nakamura}
  A.~Nakamura, T.~Saito and S.~Sakai,
  %``Lattice calculation of gluon screening masses,''
  Phys.\ Rev.\  D {\bf 69}, 014506 (2004)
  [arXiv:hep-lat/0311024].
  %%CITATION = PHRVA,D69,014506;%%

\bibitem{Wang_Guo}
    Z. X. Wang and D. R. Guo, {\it Special Functions}, World
    Scientific, Singapore (1989).


\end{thebibliography}

\end{document}